\documentclass[reprint,twocolumn]{revtex4}%
\usepackage{amssymb}
\usepackage{graphicx}
\usepackage{amsmath}
\usepackage{amssymb}
\usepackage{amscd}
\usepackage{amsfonts}%
\providecommand{\U}[1]{\protect\rule{.1in}{.1in}}
\begin{document}
\title{Time evolution of Wigner function in laser process derived by entangled state representation}
\author{{\small Li-yun Hu}$^{1,2\ast}${\small and Hong-yi Fan}$^{2}$}
\affiliation{$^{1}${\small College of Physics \& Communication Electronics, Jiangxi Normal
University, Nanchang 330022, China}}
\affiliation{$^{2}${\small Department of Physics, Shanghai Jiao Tong University, Shanghai,
200030, P.R. China}}
\affiliation{{\small *Corresponding author. E-mail addresses: hlyun2008@126.com}}

\begin{abstract}
{\small \ }Evaluating the Wigner function of quantum states in the entangled
state representation is introduced, based on which we present a new approach
for deriving time evolution formula of Wigner function in laser process.
Application of this fomula to calculating time evolution of photon number is
also presented, as an example, the case when the initial state is photon-added
coherent state is discussed.

\end{abstract}
\maketitle

\section{Introduction}

One of the major topics in Quantum Statistical Mechanics is the evolution of
pure states into mixed states \cite{1,2}. Such evolution usually happens when
a system is immersed in a thermal environment, or a signal (a quantum state)
passes through a quantum channel, and is described by a master equation.
Alternately, description of evolution of density matrices $\rho$ can be
replaced by its Wigner function's evolution in phase space \cite{3,4}. The
partial negativity of Wigner function can be considered as an indicator of
nonclassicality of quantum state. On the basis of the entangled state
representation and the thermo field dynamics we present a new approach for
deriving time evolution formula of Wigner function in amplitude-damping
channel and laser process. Application of this fomula to calculating time
evolution of photon number is also presented, as an example, the case when the
initial state is photon-added coherent state is discussed.

\section{Wigner function formula in thermo entangled state representation}

We begin with briefly reviewing the thermo entangled state representation
(TESR). On the basis of Umezawa-Takahash thermo field dynamcs
(TFD)\ \cite{5,6,7} we have constructed the TESR in doubled Fock space
\cite{8,9},
\begin{equation}
\left\vert \eta\right\rangle =\exp\left[  -\frac{1}{2}|\eta|^{2}+\eta
a^{\dagger}-\eta^{\ast}\tilde{a}^{\dagger}+a^{\dagger}\tilde{a}^{\dagger
}\right]  \left\vert 0,\tilde{0}\right\rangle , \label{2.1}%
\end{equation}
or%
\begin{equation}
\left\vert \eta\right\rangle =D\left(  \eta\right)  \left\vert \eta
=0\right\rangle ,\text{ \ }D\left(  \eta\right)  =e^{\eta a^{\dagger}%
-\eta^{\ast}a},
\end{equation}
where $D\left(  \eta\right)  $ is the displacement operator, $\tilde
{a}^{\dagger}$ is a fictitious mode accompanying the real photon creation
operator $a^{\dagger},$ $\left\vert 0,\tilde{0}\right\rangle =\left\vert
0\right\rangle \left\vert \tilde{0}\right\rangle ,$ and $\left\vert \tilde
{0}\right\rangle $ is annihilated by $\tilde{a},$ $\left[  \tilde{a},\tilde
{a}^{\dagger}\right]  =1$. \ Operating $a$ and $\tilde{a}$ on $\left\vert
\eta\right\rangle $ in Eq.(\ref{2.1}) we obtain the eigen-equations of
$\left\vert \eta\right\rangle $,%
\begin{align}
(a-\tilde{a}^{\dagger})\left\vert \eta\right\rangle  &  =\eta\left\vert
\eta\right\rangle ,\;(a^{\dagger}-\tilde{a})\left\vert \eta\right\rangle
=\eta^{\ast}\left\vert \eta\right\rangle ,\nonumber\\
\left\langle \eta\right\vert (a^{\dagger}-\tilde{a})  &  =\eta^{\ast
}\left\langle \eta\right\vert ,\ \left\langle \eta\right\vert (a-\tilde
{a}^{\dagger})=\eta\left\langle \eta\right\vert . \label{2.2}%
\end{align}
Note that $\left[  (a-\tilde{a}^{\dagger}),(a^{\dagger}-\tilde{a})\right]
=0,$ thus $\left\vert \eta\right\rangle $ is the common eigenvector of
$(a-\tilde{a}^{\dagger})$ and $(\tilde{a}-a^{\dagger}).$ Using the normally
ordered form of vacuum projector $\left\vert 0,\tilde{0}\right\rangle
\left\langle 0,\tilde{0}\right\vert =\colon\exp\left(  -a^{\dagger}a-\tilde
{a}^{\dagger}\tilde{a}\right)  \colon,$ and the technique of integration
within an ordered product (IWOP) of operators \cite{10,11,12}, we can easily
prove that $\left\vert \eta\right\rangle $ is complete and orthonormal,%
\begin{equation}
\int\frac{d^{2}\eta}{\pi}\left\vert \eta\right\rangle \left\langle
\eta\right\vert =1,\text{ }\left\langle \eta^{\prime}\right.  \left\vert
\eta\right\rangle =\pi\delta\left(  \eta^{\prime}-\eta\right)  \delta\left(
\eta^{\prime\ast}-\eta^{\ast}\right)  . \label{2.3}%
\end{equation}
It is easily seen that $\left\vert \eta=0\right\rangle $ has the properties%
\begin{equation}
\text{ \ }\left\vert \eta=0\right\rangle =e^{a^{\dagger}\tilde{a}^{\dagger}%
}\left\vert 0,\tilde{0}\right\rangle =\sum_{n=0}^{\infty}\left\vert
n,\tilde{n}\right\rangle , \label{2.4}%
\end{equation}
and%
\begin{align}
a\text{\ }\left\vert \eta=0\right\rangle  &  =\tilde{a}^{\dagger}\left\vert
\eta=0\right\rangle ,\nonumber\\
a^{\dagger}\left\vert \eta=0\right\rangle  &  =\tilde{a}\left\vert
\eta=0\right\rangle ,\label{2.5}\\
\left(  a^{\dagger}a\right)  ^{n}\left\vert \eta=0\right\rangle  &  =\left(
\tilde{a}^{\dagger}\tilde{a}\right)  ^{n}\left\vert \eta=0\right\rangle
.\nonumber
\end{align}
Note that density operators $\rho$($a^{\dagger}$,$a)$ are defined in the real
space which are commutative with operators ($\tilde{a}^{\dagger}$,$\tilde{a})$
in the tilde space.

Next, we shall derive a new expression of Wigner function in the TESR.
According to the definition of Wigner function \cite{13,14} of density
operator $\rho,$
\begin{equation}
W\left(  \alpha\right)  =\text{Tr}\left[  \Delta\left(  \alpha\right)
\rho\right]  , \label{2.6}%
\end{equation}
where $\Delta\left(  \alpha\right)  $ is the single-mode Wigner operator
\cite{13}, whose explicit form is
\begin{equation}
\Delta\left(  \alpha\right)  =\frac{1}{\pi}\colon e^{-2\left(  a^{\dagger
}-\alpha^{\ast}\right)  \left(  a-\alpha\right)  }\colon=\frac{1}{\pi}D\left(
2\alpha\right)  (-1)^{a^{\dagger}a}. \label{2.7}%
\end{equation}
By using $\left\langle \tilde{n}\right\vert \left.  \tilde{m}\right\rangle
=\delta_{n,m}$ and introducing $\left\vert \rho\right\rangle \equiv
\rho\left\vert I\right\rangle $ we can reform Eq.(\ref{2.6}) as%
\begin{align}
W\left(  \alpha\right)   &  =\sum_{m,n}^{\infty}\left\langle n,\tilde
{n}\right\vert \Delta\left(  \alpha\right)  \rho\left\vert m,\tilde
{m}\right\rangle \nonumber\\
&  =\frac{1}{\pi}\left\langle \eta=0\right\vert D\left(  2\alpha\right)
(-1)^{a^{\dagger}a}\left\vert \rho\right\rangle \nonumber\\
&  =\frac{1}{\pi}\left\langle \eta=-2\alpha\right\vert (-1)^{a^{\dagger}%
a}\left\vert \rho\right\rangle \nonumber\\
&  =\frac{1}{\pi}\left\langle \xi=2\alpha\right\vert \left.  \rho\right\rangle
, \label{2.8}%
\end{align}
where $\left\vert \xi\right\rangle $ is defined as
\begin{align}
\left\vert \xi\right\rangle _{\xi=\eta}  &  =(-1)^{a^{\dagger}a}\left\vert
\eta=-\xi\right\rangle \nonumber\\
&  =\exp\left(  -\frac{1}{2}|\xi|^{2}+\xi a^{\dagger}+\xi^{\ast}\tilde
{a}^{\dagger}-a^{\dagger}\tilde{a}^{\dagger}\right)  \left\vert 0,\tilde
{0}\right\rangle \nonumber\\
&  =D\left(  \xi\right)  e^{-a^{\dagger}\tilde{a}^{\dagger}}\left\vert
0,\tilde{0}\right\rangle . \label{2.9}%
\end{align}
It can be proved that
\begin{equation}
\left\langle \eta\right\vert \left.  \xi\right\rangle =\frac{1}{2}\exp\left(
\frac{\xi\eta^{\ast}-\xi^{\ast}\eta}{2}\right)  , \label{2.11}%
\end{equation}
a Fourier transformation kernel, so $\left\vert \xi\right\rangle $ can be
considered the conjugate state of $\left\vert \eta\right\rangle ,$ which also
possess orthonormal and complete properties%
\begin{equation}
\int\frac{d^{2}\xi}{\pi}\left\vert \xi\right\rangle \left\langle
\xi\right\vert =1,\text{ }\left\langle \xi^{\prime}\right.  \left\vert
\xi\right\rangle =\pi\delta\left(  \xi^{\prime}-\xi\right)  \delta\left(
\xi^{\prime\ast}-\xi^{\ast}\right)  . \label{2.10}%
\end{equation}
Eq.(\ref{2.8}) is just a new formula for evaluating the Wigner function of
quantum states: by calculating the overlap between two \textquotedblleft pure
states\textquotedblright\ in enlarged Fock space rather than using the
ensemble average in real mode space.

For example, for number state $\left\vert n\right\rangle \left\langle
n\right\vert ,$ noticing $\left\vert n\right\rangle \left\langle n\right\vert
\left.  I\right\rangle =\left\vert n,\tilde{n}\right\rangle $, and the
generating function of two-variable Hermite polynomial \cite{15,16}
$H_{m,n}\left(  x,y\right)  $,%
\begin{equation}
\sum_{m,n}^{\infty}\frac{t^{m}t^{\prime n}}{m!n!}H_{m,n}\left(  x,y\right)
=\exp\left[  -tt^{\prime}+tx+t^{\prime}y\right]  , \label{2.12}%
\end{equation}
we see
\begin{align}
W_{\left\vert n\right\rangle \left\langle n\right\vert }\left(  \alpha\right)
&  =\frac{1}{\pi}\left\langle n,\tilde{n}\right\vert \left.  \xi_{=2\alpha
}\right\rangle =\frac{1}{n!\pi}e^{-\frac{1}{2}|\xi|^{2}}H_{n,n}\left(  \xi
,\xi^{\ast}\right) \nonumber\\
&  =\frac{\left(  -1\right)  ^{n}}{\pi}e^{-2|\alpha|^{2}}L_{n}\left(
4|\alpha|^{2}\right)  , \label{2.13}%
\end{align}
in the last step in Eq.(\ref{2.13}) we have used the relation between
$H_{m,n}\left(  x,y\right)  $ and Laguerre polynomial $L_{m}\left(  x\right)
$ \cite{17},
\begin{equation}
L_{n}\left(  xy\right)  =\frac{\left(  -1\right)  ^{n}}{n!}H_{n,n}\left(
x,y\right)  . \label{2.14}%
\end{equation}
Similarly, for coherent state $\left\vert z\right\rangle \left\langle
z\right\vert $ ($\left\vert z\right\rangle =\exp(-\left\vert z\right\vert
^{2}/2+za^{\dagger})\left\vert 0\right\rangle $) \cite{18,19}, due to
$\left\vert z\right\rangle \left\langle z\right\vert \left.  I\right\rangle
=D\left(  z\right)  \tilde{D}\left(  z^{\ast}\right)  \left\vert 0\tilde
{0}\right\rangle =\left\vert z,\tilde{z}^{\ast}\right\rangle ,$ we have
\begin{align}
W_{\left\vert z\right\rangle \left\langle z\right\vert }\left(  \alpha\right)
&  =\frac{1}{\pi}\left\langle 0,\tilde{0}\right\vert \exp\left(
-2|\alpha|^{2}+2\alpha^{\ast}a^{\dagger}+2\alpha\tilde{a}-a\tilde{a}\right)
\left\vert z,\tilde{z}^{\ast}\right\rangle \nonumber\\
&  =\frac{1}{\pi}\exp\left[  -2\left\vert \alpha-z\right\vert ^{2}\right]  .
\label{2.15}%
\end{align}
Further, using Eq.(\ref{2.11}) and the completeness of $\left\langle
\eta\right\vert $ in Eq.(\ref{2.3}), we can reform Eq.(\ref{2.8}) as%
\begin{equation}
W\left(  \alpha\right)  =\int\frac{d^{2}\eta}{\pi^{2}}\left\langle \xi
=2\alpha\right\vert \left.  \eta\right\rangle \left\langle \eta\right\vert
\left.  \rho\right\rangle =\int\frac{d^{2}\eta}{2\pi^{2}}e^{\alpha^{\ast}%
\eta-\alpha\eta^{\ast}}\left\langle \eta\right\vert \left.  \rho\right\rangle
. \label{2.16}%
\end{equation}
Once $\left\langle \eta\right\vert \left.  \rho\right\rangle $ is known, one
can calculate the Wigner function by taking the Fourier transform of
$\left\langle \eta\right\vert \left.  \rho\right\rangle $. Eqs. (\ref{2.8})
and (\ref{2.12}) are two ways accessing to Wigner function, we can use either
one to derive Wigner functions.

\section{Evolution formula of Wigner function for amplitude damping channel}

In this section, we consider Wigner function's time evolution in the amplitude
decay channel (dissipation in a lossy cavity) described by the following
master equation \cite{20}%

\begin{equation}
\frac{d\rho}{dt}=\kappa\left(  2a\rho a^{\dagger}-a^{\dagger}a\rho-\rho
a^{\dagger}a\right)  , \label{3.1}%
\end{equation}
where $\kappa$ is the rate of decay. In Ref. \cite{21} we have reformed
(\ref{3.1}) as\
\begin{equation}
\frac{d}{dt}\left\vert \rho\right\rangle =\kappa\left(  2a\tilde{a}%
-a^{\dagger}a-\tilde{a}^{\dagger}\tilde{a}\right)  \left\vert \rho
\right\rangle , \label{3.2}%
\end{equation}
thus the formal solution of Eq.(\ref{3.2}) is
\begin{equation}
\left\vert \rho\left(  t\right)  \right\rangle =e^{\kappa t\left(  a\tilde
{a}-\tilde{a}^{\dagger}a^{\dagger}+1\right)  }e^{\left(  1-e^{2\kappa
t}\right)  \left(  a^{\dagger}-\tilde{a}\right)  \left(  a-\tilde{a}^{\dagger
}\right)  /2}\left\vert \rho_{0}\right\rangle . \label{3.7}%
\end{equation}
Then projecting Eq.(\ref{3.7}) on $\left\langle \eta\right\vert $, and
noticing $\exp\left[  \kappa t\left(  a\tilde{a}-\tilde{a}^{\dagger}%
a^{\dagger}\right)  \right]  $ being the two-mode squeezing operator,%
\begin{equation}
\left\langle \eta\right\vert \exp\left[  \kappa t\left(  a\tilde{a}-\tilde
{a}^{\dagger}a^{\dagger}\right)  \right]  =e^{-\kappa t}\left\langle \eta
e^{-\kappa t}\right\vert , \label{3.8}%
\end{equation}
as well as Eq.(\ref{2.2}), we obtain%
\begin{equation}
\left\langle \eta\right.  \left\vert \rho\left(  t\right)  \right\rangle
=e^{-\frac{1}{2}T\left\vert \eta\right\vert ^{2}}\left\langle \eta e^{-\kappa
t}\right\vert \left.  \rho_{0}\right\rangle , \label{3.9}%
\end{equation}
where $T=1-e^{-2\kappa t}.$ Substituting Eq.(\ref{3.9}) into Eq.(\ref{2.16}),
we derive the Wigner function at time $t$%
\begin{equation}
W\left(  \alpha,t\right)  =\int\frac{d^{2}\eta}{2\pi^{2}}e^{\alpha^{\ast}%
\eta-\alpha\eta^{\ast}-\frac{1}{2}T\left\vert \eta\right\vert ^{2}%
}\left\langle \eta e^{-\kappa t}\right\vert \left.  \rho_{0}\right\rangle .
\label{3.10}%
\end{equation}
Inserting the completeness relation (\ref{2.10}) into Eq.(\ref{3.10}) and
noticing Eqs.(\ref{2.8}) as well as (\ref{2.11}), we can reform Eq.(\ref{3.10}%
) as%
\begin{align}
W\left(  \alpha,t\right)   &  =\int\frac{d^{2}\xi^{\prime}}{\pi}\int
\frac{d^{2}\eta}{2\pi^{2}}e^{-\frac{1}{2}T\left\vert \eta\right\vert ^{2}%
}\left\langle \eta e^{-\kappa t}\right.  \left\vert \xi^{\prime}\right\rangle
\left\langle \xi^{\prime}\right\vert \left.  \rho_{0}\right\rangle \nonumber\\
&  =\int\frac{d^{2}\beta d^{2}\eta}{\pi^{2}}e^{-\frac{T}{2}\left\vert
\eta\right\vert ^{2}+\eta\left(  \alpha^{\ast}-\beta^{\ast}e^{-\kappa
t}\right)  +\eta^{\ast}\left(  \beta e^{-\kappa t}-\alpha\right)  }W\left(
\beta,0\right) \nonumber\\
&  =\frac{2}{T}\int\frac{d^{2}\beta}{\pi}\exp\left[  -\frac{2}{T}\left\vert
\alpha-\beta e^{-\kappa t}\right\vert ^{2}\right]  W\left(  \beta,0\right)
\label{3.11}%
\end{align}
where $W\left(  \beta,0\right)  $ is the Wigner function at initial time, and
we have used the following integral formula \cite{17}
\begin{equation}
\int\frac{d^{2}z}{\pi}\exp\left(  \zeta\left\vert z\right\vert ^{2}+\xi z+\eta
z^{\ast}\right)  =-\frac{1}{\zeta}e^{-\frac{\xi\eta}{\zeta}},\text{Re}\left(
\zeta\right)  <0. \label{3.12}%
\end{equation}
Eq.(\ref{3.11}) is the expression of time evolution of Wigner function for
amplitude damping channel.

For example, for the photon-added coherent state $C_{m}a^{\dag m}\left\vert
z\right\rangle $, where $C_{m}=[m!L_{m}(-\left\vert z\right\vert ^{2})]^{-1}$
is the normalization factor, the initial Wigner function $W\left(
\beta,0\right)  $ is given by \cite{22}%
\begin{equation}
W\left(  \beta,0\right)  =\frac{\left(  -1\right)  ^{m}e^{-2\left\vert
\beta-z\right\vert ^{2}}}{\pi L_{m}(-\left\vert z\right\vert ^{2})}%
L_{m}(\left\vert 2\beta-z\right\vert ^{2}). \label{3.15}%
\end{equation}
Substituting Eq.(\ref{3.15}) into Eq.(\ref{3.11}) and using Eq.(\ref{2.14}) as
well as the another generating function of $H_{m,n}\left(  x,y\right)  ,$
\begin{equation}
H_{m,n}\left(  x,y\right)  =\left.  \frac{\partial^{m+n}}{\partial\tau
^{m}\partial\tau^{\prime n}}\exp\left[  -\tau\tau^{\prime}+\tau x+\tau
^{\prime}y\right]  \right\vert _{\tau=\tau^{\prime}=0}, \label{3.16}%
\end{equation}
we have%
\begin{align}
W\left(  \alpha,t\right)   &  =\frac{2}{T}\frac{e^{-2\left(  \left\vert
z\right\vert ^{2}+\frac{1}{T}\left\vert \alpha\right\vert ^{2}\right)  }}{\pi
m!L_{m}(-\left\vert z\right\vert ^{2})}\frac{\partial^{2m}}{\partial\tau
^{m}\partial\tau^{\prime m}}e^{-\tau\tau^{\prime}-\tau z-z^{\ast}\tau^{\prime
}}\nonumber\\
&  \int\frac{d^{2}\beta}{\pi}\exp\left[  -\frac{\allowbreak2}{T}\left\vert
\beta\right\vert ^{2}+2\beta\left(  z^{\ast}+\allowbreak\frac{\alpha^{\ast}%
}{T}e^{-t\kappa}+\tau\right)  \right. \nonumber\\
&  \left.  +2\beta^{\ast}\left(  z+\frac{\alpha}{T}e^{-t\kappa}+\tau^{\prime
}\right)  \right]  _{\tau=\tau^{\prime}=0}\nonumber\\
&  =\frac{e^{-2\allowbreak\left\vert \alpha-ze^{-\kappa t}\right\vert ^{2}}%
}{\pi m!L_{m}(-\left\vert z\right\vert ^{2})}\frac{\partial^{2m}}{\partial
\tau^{m}\partial\tau^{\prime m}}\exp\left[  \left(  1-2e^{-2t\kappa}\right)
\tau\tau^{\prime}\right. \nonumber\\
&  +\left[  \left(  1-2e^{-2t\kappa}\right)  z+2\alpha e^{-t\kappa}\right]
\tau\nonumber\\
&  +\left.  \left[  \allowbreak\left(  1-2e^{-2t\kappa}\right)  z^{\ast
}+2\alpha^{\ast}e^{-t\kappa}\right]  \tau^{\prime}\right]  _{\tau=\tau
^{\prime}=0}. \label{3.17}%
\end{align}
With use of a scaled transformation in the right-hand part of Eq.(\ref{3.17})
we finally get%

\begin{align}
W\left(  \alpha,t\right)   &  =\frac{\left(  1-2e^{-2\kappa t}\right)  ^{m}%
}{\pi L_{m}(-\left\vert z\right\vert ^{2})}e^{-2\allowbreak\left\vert
\alpha-ze^{-\kappa t}\right\vert ^{2}}\nonumber\\
&  \times L_{m}\left[  -\frac{\left\vert 2\alpha e^{-\kappa t}+z\left(
1-2e^{-2\kappa t}\right)  \right\vert ^{2}}{1-2e^{-2\kappa t}}\right]  ,
\label{3.18}%
\end{align}
which is the analytical expression of the time evolution of Wigner function
for any number ($m$) photon-added coherent state in photon loss channel
\cite{23}. In particular, when $t=0,$ Eq.(\ref{3.18}) just reduce to
Eq.(\ref{3.15}).

\section{Evolution formula of Wigner function for Laser process}

We now generalize the master equation to the case of Laser theory. The
mechanism of laser is described by the following master equation%
\begin{align}
\frac{d\rho\left(  t\right)  }{dt}  &  =g\left[  2a^{\dagger}\rho\left(
t\right)  a-aa^{\dagger}\rho\left(  t\right)  -\rho\left(  t\right)
aa^{\dagger}\right] \nonumber\\
&  +\kappa\left[  2a\rho\left(  t\right)  a^{\dagger}-a^{\dagger}a\rho\left(
t\right)  -\rho\left(  t\right)  a^{\dagger}a\right]  , \label{4.1}%
\end{align}
where $g$\ and $\kappa$\ are the cavity gain and the loss, respectively.
Eq.(\ref{4.1}) reduces to Eq.(\ref{3.1}) when $g=0;$\ while for $g\rightarrow
\kappa\bar{n}$ and $\kappa\rightarrow\kappa\left(  \bar{n}+1\right)  ,$
Eq.(\ref{4.1}) becomes
\begin{align}
\frac{d\rho}{dt}  &  =\kappa\left(  \bar{n}+1\right)  \left(  2a\rho
a^{\dagger}-a^{\dagger}a\rho-\rho a^{\dagger}a\right) \nonumber\\
&  +\kappa\bar{n}\left(  2a^{\dagger}\rho a-aa^{\dagger}\rho-\rho aa^{\dagger
}\right)  , \label{4.2}%
\end{align}
which corresponds to the master equation in thermal environment \cite{20}.

Similar to the way of deriving Eq.(\ref{3.9}), we have derived in Ref.
\cite{21}%
\begin{align}
\left\vert \rho\left(  t\right)  \right\rangle  &  =\exp\left[  \left(
a\tilde{a}-\tilde{a}^{\dagger}a^{\dagger}+1\right)  \left(  \kappa-g\right)
t\right] \nonumber\\
&  \times\exp\left[  \frac{\left(  \kappa+g\right)  \left(  1-e^{2\left(
\kappa-g\right)  t}\right)  }{2\left(  \kappa-g\right)  }\left(  a^{\dagger
}-\tilde{a}\right)  \left(  a-\tilde{a}^{\dagger}\right)  \right]  \left\vert
\rho_{0}\right\rangle . \label{4.6}%
\end{align}
\ Thus the matrix element $\left\langle \eta\right\vert \left.  \rho\left(
t\right)  \right\rangle $\ is given by%
\begin{equation}
\left\langle \eta\right\vert \left.  \rho\left(  t\right)  \right\rangle
=\exp\left[  -\frac{A}{2}|\eta|^{2}\right]  \left\langle \eta e^{-\left(
\kappa-g\right)  t}\right\vert \left.  \rho_{0}\right\rangle , \label{4.7}%
\end{equation}
where
\begin{equation}
A=\frac{\kappa+g}{\kappa-g}\left(  1-e^{-2\left(  \kappa-g\right)  t}\right)
. \label{4.9}%
\end{equation}
According to Eq.(\ref{2.12}) the Wigner function's evolution for Laser process
is given by%
\begin{align}
&  W\left(  \alpha,t\right) \nonumber\\
&  =\int\frac{d^{2}\eta}{2\pi^{2}}e^{-\frac{A}{2}|\eta|^{2}+\alpha^{\ast}%
\eta-\alpha\eta^{\ast}}\left\langle \eta e^{-\left(  \kappa-g\right)
t}\right\vert \left.  \rho_{0}\right\rangle \nonumber\\
&  =\int\frac{d^{2}\xi d^{2}\eta}{2\pi^{2}}e^{-\frac{A}{2}|\eta|^{2}%
+\alpha^{\ast}\eta-\alpha\eta^{\ast}}\left\langle \eta e^{-\left(
\kappa-g\right)  t}\right.  \left\vert \xi_{=2\beta}\right\rangle W\left(
\beta,0\right) \nonumber\\
&  =\int\frac{d^{2}\xi d^{2}\eta}{\pi^{2}}e^{-\frac{A}{2}|\eta|^{2}%
+\eta\left(  \alpha^{\ast}-\beta^{\ast}e^{-\left(  \kappa-g\right)  t}\right)
+\eta^{\ast}\left(  \beta e^{-\left(  \kappa-g\right)  t}-\alpha\right)
}W\left(  \beta,0\right) \nonumber\\
&  =\frac{2}{A}\int\frac{d^{2}\beta}{\pi}\exp\left[  -\frac{2}{A}\left\vert
\alpha-\beta e^{-\left(  \kappa-g\right)  t}\right\vert ^{2}\right]  W\left(
\beta,0\right)  , \label{4.8}%
\end{align}
where we have used Eq.(\ref{3.12}). In particular, when $g=0,$ Eq.(\ref{4.8})
reduces to Eq.(\ref{3.11}). For $g\rightarrow\kappa\bar{n}$ and $\kappa
\rightarrow\kappa\left(  \bar{n}+1\right)  $, leading to $A=\left(  2\bar
{n}+1\right)  T,$ Eq.(\ref{4.8}) becomes%
\begin{equation}
W\left(  \alpha,t\right)  =\frac{2}{\left(  2\bar{n}+1\right)  T}\int
\frac{d^{2}\beta}{\pi}W\left(  \beta,0\right)  e^{-2\frac{\allowbreak
\left\vert \alpha-\beta e^{-\kappa t}\right\vert ^{2}}{\left(  2\allowbreak
\bar{n}+1\right)  T}}, \label{4.10}%
\end{equation}
or
\begin{equation}
W\left(  \alpha,t\right)  =2e^{2\kappa t}\int d^{2}\beta W_{T}\left(
\beta\right)  W\left(  e^{\kappa t}(\alpha-\sqrt{T}\beta),0\right)  ,
\label{4.11}%
\end{equation}
where $W_{T}\left(  \beta\right)  =\frac{1}{\pi\left(  2\bar{n}+1\right)
}e^{-\frac{2\allowbreak\left\vert \beta\right\vert ^{2}}{2\allowbreak\bar
{n}+1}}$ is the Wigner function of the thermal state\ with mean photon number
$\bar{n}$.

Similar to the way of deriving Eq.(\ref{3.18}), when the initial state is
$C_{m}a^{\dag m}\left\vert z\right\rangle ,$ substituting Eq.(\ref{3.15}) into
Eq.(\ref{4.8}) we have
\begin{align}
W\left(  \beta,\beta^{\ast},t\right)    & =\frac{e^{-C-2\left\vert
\beta\right\vert ^{2}}}{\pi L_{m}(-\left\vert z\right\vert ^{2})}\frac{A^{m}%
}{\left(  2\bar{n}T+1\right)  }\nonumber\\
& \times\frac{\left[  \left(  \bar{n}+1\right)  T\right]  ^{m}}{\left(
\bar{n}T+1\right)  ^{m}}L_{m}\left(  -\frac{\left\vert B\right\vert ^{2}}%
{A}\right)  ,\label{4.12}%
\end{align}
where%
\begin{align}
A &  =1-\allowbreak\frac{e^{-2\kappa t}/T}{\left(  2T\bar{n}+1\right)  \left(
\bar{n}+1\right)  },\nonumber\\
B &  =\sqrt{\frac{\left(  \bar{n}+1\right)  T}{\bar{n}T+1}}z^{\ast}%
+\frac{\sqrt{\bar{n}T+1}e^{-\kappa t}\left(  2\beta^{\ast}-\frac{z^{\ast
}e^{-\kappa t}}{\bar{n}T+1}\right)  }{\left(  2\bar{n}T+1\right)
\sqrt{\left(  \bar{n}+1\right)  T}},\nonumber\\
C &  =\allowbreak\frac{1}{2\bar{n}T+1}\left(  \frac{3\bar{n}T+2}{T\bar{n}%
+1}\left\vert ze^{-\kappa t}\right\vert ^{2}+4T^{2}\bar{n}^{2}\left\vert
\beta\right\vert ^{2}\right)  \nonumber\\
&  \text{ \ \ }-\allowbreak\frac{2e^{-\kappa t}\left(  T\bar{n}+1\right)
}{2\bar{n}T+1}\allowbreak\left(  z\beta^{\ast}+\beta z^{\ast}\right)
.\label{4.13}%
\end{align}
In particular, when $\bar{n}=0,$ leading to $A=\allowbreak\frac
{1-2e^{-2t\kappa}}{T},B=\frac{1}{\sqrt{T}}\left(  \left(  1-2e^{-2\kappa
t}\right)  z^{\ast}+2e^{-\kappa t}\beta^{\ast}\right)  ,$ and $-C-2\left\vert
\beta\right\vert ^{2}=-2\left\vert \beta-ze^{-t\kappa}\right\vert ^{2},$ thus
Eq.(\ref{4.13}) reduces to Eq.(\ref{3.18}).

Eq.(\ref{4.12}) manifestly shows that the Wigner function of $C_{m}a^{\dag
m}\left\vert z\right\rangle $ in thermal environment is closely related to the
Laguerre polynomials. In addition, due to $L_{m}\left(  -\left\vert
x\right\vert ^{2}\right)  >0,$ so $C_{m}>0,$ thus it is easily seen that when
$A>0,$ which means the condition
\begin{equation}
\kappa t\geqslant\kappa t_{c}=\frac{1}{2}\ln\frac{2\left(  \bar{n}+1\right)
}{2\bar{n}+1}, \label{4.14}%
\end{equation}
the Wigner function (\ref{4.12}) is always positive-definite. Thus we
emphasize that for any values of $m$, when the condition (\ref{4.14}) is
satisfied, the Wigner function has no chance to be negative.

\section{Time evolution of photon number for the laser process}

Next we consider the photon number (PN) of\ density operator $\rho$ for the
laser process. According to the TFD, we can reform the PN $p\left(  n\right)
=\left\langle n\right\vert \rho\left\vert n\right\rangle $ as
\begin{align}
p\left(  n\right)   &  =\left\langle n\right\vert \rho\left\vert
n\right\rangle =\sum_{m=0}^{\infty}\left\langle n,\tilde{n}\right\vert
\rho\left\vert m,\tilde{m}\right\rangle \nonumber\\
&  =\left\langle n,\tilde{n}\right\vert \rho\left\vert I\right\rangle
=\left\langle n,\tilde{n}\right\vert \left.  \rho\right\rangle , \label{5.1}%
\end{align}
thus the PN is converted to the matrix element $\left\langle n,\tilde
{n}\right\vert \left.  \rho\right\rangle $ in thermo dynamics frame. Then
using the completeness of $\left\langle \xi\right\vert $ and Eq.(\ref{2.8}) as
well as Eq.(\ref{2.13}), we see%
\begin{align}
p\left(  n\right)   &  =\int\frac{d^{2}\xi}{\pi}\left\langle n,\tilde
{n}\right\vert \left.  \xi\right\rangle \left\langle \xi\right\vert \left.
\rho\right\rangle \nonumber\\
&  =\int d^{2}\xi\left\langle n,\tilde{n}\right\vert \left.  \xi\right\rangle
W\left(  \alpha=\xi/2\right) \nonumber\\
&  =4\pi\int d^{2}\alpha W_{\left\vert n\right\rangle \left\langle
n\right\vert }\left(  \alpha\right)  W\left(  \alpha\right)  , \label{5.2}%
\end{align}
one can see this formula also in \cite{1,24}. Thus one can calculate the PN by
combining Eq.(\ref{4.8}) and (\ref{5.2}).

Now we evaluate the PN of the above decoherence model in Eq.(\ref{4.1}).
Substituting Eq.(\ref{4.8}) into Eq.(\ref{5.2}), we see%
\begin{equation}
p\left(  n\right)  =\frac{8}{A}\int d^{2}\beta W\left(  \beta,0\right)
G\left(  \beta\right)  , \label{5.3}%
\end{equation}
where
\begin{align}
G\left(  \beta\right)   &  \equiv\int d^{2}\alpha W_{\left\vert n\right\rangle
\left\langle n\right\vert }\left(  \alpha\right)  e^{-\frac{2}{A}\left\vert
\alpha-\beta e^{-\left(  \kappa-g\right)  t}\right\vert ^{2}-2|\alpha|^{2}%
}\nonumber\\
&  =(-1)^{n}\int\frac{d^{2}\alpha}{\pi}L_{n}\left(  4\left\vert \alpha
\right\vert ^{2}\right)  e^{-\frac{2}{A}\left\vert \alpha-\beta e^{-\left(
\kappa-g\right)  t}\right\vert ^{2}-2|\alpha|^{2}}. \label{5.4}%
\end{align}
Using Eqs.(\ref{3.12}) and (\ref{3.16}) we can evaluate Eq.(\ref{5.4}) as%
\begin{align}
G\left(  \beta\right)   &  =\frac{1}{n!}\frac{\partial^{n+n}}{\partial\tau
^{n}\partial\tau^{\prime n}}e^{-\tau\tau^{\prime}-\frac{2}{A}\left\vert
\beta\right\vert ^{2}e^{-2\left(  \kappa-g\right)  t}}\int\frac{d^{2}\alpha
}{\pi}\nonumber\\
&  \times\exp\left[  -2\frac{A+1}{A}|\alpha|^{2}+2\alpha\left(  \tau
+\frac{\beta^{\ast}}{Ae^{\left(  \kappa-g\right)  t}}\right)  \right.
\nonumber\\
&  \left.  +2\alpha^{\ast}\left(  \tau^{\prime}+\frac{\beta}{Ae^{\left(
\kappa-g\right)  t}}\right)  \right]  _{\tau=\tau^{\prime}=0}\nonumber\\
&  =\frac{Ae^{-\frac{2e^{-2\left(  \kappa-g\right)  t}}{A+1}\left\vert
\beta\right\vert ^{2}}}{2\left(  A+1\right)  n!}\frac{\partial^{n+n}}%
{\partial\tau^{n}\partial\tau^{\prime n}}\exp\left[  -\frac{1-A}{1+A}%
\tau^{\prime}\tau\right. \nonumber\\
&  +\left.  \frac{2\beta^{\ast}e^{-\left(  \kappa-g\right)  t}}{A+1}%
\tau^{\prime}+\tau\frac{2\beta e^{-\left(  \kappa-g\right)  t}}{A+1}\right]
_{\tau=\tau^{\prime}=0}. \label{5.5}%
\end{align}
After making some scaled transformations, we finally obtain%
\begin{align}
G\left(  \beta\right)   &  =\frac{A\left(  A-1\right)  ^{n}}{2\left(
1+A\right)  ^{n+1}n!}e^{-\frac{2e^{-2\left(  \kappa-g\right)  t}}%
{A+1}\left\vert \beta\right\vert ^{2}}\nonumber\\
&  \times\frac{\partial^{n+n}}{\partial\tau^{n}\partial\tau^{\prime n}}\left.
e^{-\tau^{\prime}\tau+\frac{2\beta^{\ast}e^{-\left(  \kappa-g\right)  t}%
}{\sqrt{1-A^{2}}}\tau^{\prime}+\tau\frac{2\beta e^{-\left(  \kappa-g\right)
t}}{\sqrt{1-A^{2}}}}\right\vert _{\tau=\tau^{\prime}=0}\nonumber\\
&  =\frac{A\left(  A-1\right)  ^{n}}{2\left(  1+A\right)  ^{n+1}}%
e^{-\frac{2e^{-2\left(  \kappa-g\right)  t}}{A+1}\left\vert \beta\right\vert
^{2}}L_{n}\left(  \frac{4\left\vert \beta\right\vert ^{2}e^{-2\left(
\kappa-g\right)  t}}{1-A^{2}}\right)  . \label{5.6}%
\end{align}
Substituting Eq.(\ref{5.6}) into Eq.(\ref{5.3}) yields%
\begin{align}
p\left(  n\right)   &  =\frac{4\left(  A-1\right)  ^{n}}{\left(  A+1\right)
^{n+1}}\int d^{2}\beta e^{-\frac{2e^{-2\left(  \kappa-g\right)  t}}%
{A+1}\left\vert \beta\right\vert ^{2}}\nonumber\\
&  \times L_{n}\left\{  \frac{4e^{-2\left(  \kappa-g\right)  t}}{1-A^{2}%
}\left\vert \beta\right\vert ^{2}\right\}  W\left(  \beta,0\right)  ,
\label{5.7}%
\end{align}
which is a new formula for calculating the photon number distribution of the
open system in enviornment. From Eq.(\ref{5.7}) it is easily seen that once
the Wigner function of initial state is known, one can obtain its photon
number distribution by performing the integration in Eq.(\ref{5.7}).

In particular, when $g=0,$ $A=1-e^{-2\kappa t}=T,$ Eq.(\ref{5.7}) reduces to
\begin{align}
p\left(  n\right)   &  =\frac{4(-1)^{n}e^{2\kappa t}}{\left(  2e^{2\kappa
t}-1\right)  ^{n+1}}\int d^{2}\beta e^{-\frac{2}{2e^{2\kappa t}-1}\left\vert
\beta\right\vert ^{2}}\nonumber\\
&  \times L_{n}\left\{  \frac{4e^{2\kappa t}}{2e^{2\kappa t}-1}\left\vert
\beta\right\vert ^{2}\right\}  W\left(  \beta,0\right)  , \label{5.8}%
\end{align}
which corresponds to the photon number of density operator in the amplitude
damping quantum channel.

While for $g\rightarrow\kappa\bar{n}$ and $\kappa\rightarrow\kappa\left(
\bar{n}+1\right)  $, $,$ Eq.(\ref{5.7}) becomes to%
\begin{align}
p\left(  n\right)   &  =\frac{4\left(  \mathcal{A}-1\right)  ^{n}}{\left(
\mathcal{A}+1\right)  ^{n+1}}\int d^{2}\beta e^{-\frac{2e^{-2\kappa t}%
}{\mathcal{A}+1}\left\vert \beta\right\vert ^{2}}\nonumber\\
&  \times L_{n}\left\{  \frac{4e^{-2\kappa t}}{1-\mathcal{A}^{2}}\left\vert
\beta\right\vert ^{2}\right\}  W\left(  \beta,0\right)  , \label{5.9}%
\end{align}
where $\mathcal{A}=\left(  2\bar{n}+1\right)  T=\left(  2\bar{n}+1\right)
\left(  1-e^{-2\kappa t}\right)  .$ Eq.(\ref{5.9}) corresponds to the photon
number of system interacting with thermal bath.

For example, we still consider the photon-added coherent state field.
Substituting Eq.(\ref{3.15}) into Eq.(\ref{5.7}) and uisng Eqs.(\ref{3.12})
and (\ref{3.16}) yields%
\begin{align}
p\left(  n\right)   &  =Ne^{-2\left\vert z\right\vert ^{2}}\int\frac
{d^{2}\beta}{\pi}L_{m}(\left\vert 2\beta-z\right\vert ^{2})L_{n}\left\{
\frac{4e^{-2\left(  \kappa-g\right)  t}}{1-A^{2}}\left\vert \beta\right\vert
^{2}\right\}  \nonumber\\
&  \times\exp\left[  \allowbreak2\left(  z\beta^{\ast}+\beta z^{\ast}\right)
-2\left(  \allowbreak1+\frac{e^{-2\left(  \kappa-g\right)  t}}{A+1}\right)
\left\vert \beta\right\vert ^{2}\right]  \nonumber\\
&  =\frac{Ne^{-2\left\vert z\right\vert ^{2}}\left(  -1\right)  ^{m+n}}%
{m!n!}\frac{\partial^{2m}}{\partial\upsilon^{m}\partial\upsilon^{\prime m}%
}\frac{\partial^{2n}}{\partial\tau^{n}\partial\tau^{\prime n}}e^{-\upsilon
\upsilon^{\prime}-z^{\ast}\upsilon^{\prime}}\nonumber\\
&  \times e^{-z\upsilon-\tau\tau^{\prime}}\int\frac{d^{2}\beta}{\pi}%
\exp\left[  -2\mu\left\vert \beta\right\vert ^{2}+2\left(  \sigma\tau+z^{\ast
}+\upsilon\right)  \beta\right.  \nonumber\\
&  \left.  +2\left(  \sigma\tau^{\prime}+z+\upsilon^{\prime}\right)
\beta^{\ast}\right]  _{\upsilon=\upsilon^{\prime}=\tau=\tau^{\prime}%
=0}\nonumber\\
&  =\frac{\allowbreak N\left(  -1\right)  ^{m+n}}{2\mu m!n!}e^{\frac{2-2\mu
}{\mu}\left\vert z\right\vert ^{2}}\frac{\partial^{2m}}{\partial\upsilon
^{m}\partial\upsilon^{\prime m}}\frac{\partial^{2n}}{\partial\tau^{n}%
\partial\tau^{\prime n}}\nonumber\\
&  \times\exp\left[  \omega\upsilon\upsilon^{\prime}+\allowbreak\left(
\lambda\sigma-1\right)  \allowbreak\tau\tau^{\prime}+\lambda\left(
\tau\upsilon^{\prime}+\upsilon\tau^{\prime}\right)  \right.  \nonumber\\
&  \left.  +\omega\left(  z^{\ast}\upsilon^{\prime}+z\upsilon\right)
+\lambda\left(  z\tau+z^{\ast}\tau^{\prime}\right)  \right]  _{\upsilon
=\upsilon^{\prime}=\tau=\tau^{\prime}=0},\label{5.10}%
\end{align}
where we have set%
\begin{equation}
\omega=\frac{2-\mu}{\mu},\lambda=\frac{2\sigma}{\mu},\text{ }\sigma
=\frac{e^{-\left(  \kappa-g\right)  t}}{\sqrt{1-A^{2}}},\label{5.11}%
\end{equation}
and%
\begin{equation}
N=\frac{4\left(  A-1\right)  ^{n}}{\left(  A+1\right)  ^{n+1}}\frac{\left(
-1\right)  ^{m}}{L_{m}(-\left\vert z\right\vert ^{2})},\mu=\allowbreak
1+\frac{e^{-2\left(  \kappa-g\right)  t}}{A+1}.\label{5.12}%
\end{equation}
Further expanding the exponential item $\exp\left[  \omega\upsilon
\upsilon^{\prime}+\allowbreak\left(  \lambda\sigma-1\right)  \tau\tau^{\prime
}\right]  ,$ we finally obtain%
\begin{align}
p\left(  n\right)   &  =\frac{\allowbreak N\lambda^{2n}e^{\frac{2-2\mu}{\mu
}\left\vert z\right\vert ^{2}}}{2\mu\left(  -\omega\right)  ^{n-m}}%
\sum_{l,k=0}^{m,n}\frac{m!n!\left[  \omega\left(  \lambda\sigma-1\right)
/\lambda^{2}\right]  ^{k}}{l!k!\left[  \left(  m-l\right)  !\left(
n-k\right)  !\right]  ^{2}}\nonumber\\
&  \times\left\vert H_{m-l,n-k}\left(  i\sqrt{\omega}z,i\sqrt{\omega}z^{\ast
}\right)  \right\vert ^{2}.\label{5.13}%
\end{align}
In particular, when $g=0,$ leading to $A=\omega=T,\sigma=\frac{e^{-\kappa t}%
}{\sqrt{1-T^{2}}},\mu=\allowbreak\frac{2}{2-e^{-2\kappa t}},\lambda\sigma=1,$
and $\lambda=\sqrt{1+T}$, thus%
\begin{align}
p\left(  n\right)   &  =\frac{m!}{n!}\frac{\left(  1-\omega\right)  ^{n}%
}{L_{m}(-\left\vert z\right\vert ^{2})}\sum_{l=0}^{m}\frac{\omega
^{m-n}e^{-e^{-2t\kappa}\left\vert z\right\vert ^{2}}}{l!\left[  \left(
m-l\right)  !\right]  ^{2}}\nonumber\\
&  \times\left\vert H_{m-l,n}\left(  i\sqrt{\omega}z,i\sqrt{\omega}z^{\ast
}\right)  \right\vert ^{2},\label{5.14}%
\end{align}
which concides with Eq.(43) with idea detection efficiency in Ref. \cite{23}.

In sum, by virtue of the thermo entangled state representation that has a
fictitious mode as a counterpart mode of the system mode, we have derived the
relation between the Wigner functions at $t$ time and the initial time when
quantum system interacts with envoirnment, such as decoherence, damping and
amplification. As another quantity describing quantum system, the formula of
photon number distribution has also been derived, which can be evaluated by
performing an integration for the initial Wigner function. Our deriviations
seem more concise.

\textbf{ACKNOWLEDGEMENT:} Work supported by the National Natural Science
Foundation of China under grants: 10775097 and 10874174.


\begin{thebibliography}{99}                                                                                               %


\bibitem {1}W. H. Louisell, \textit{Quantum Statistical Properties of
Radiation} (Wiley, New York, 1973).

\bibitem {2}H. J. Carmichael, \textit{Statistical Methods in Quantum Optics 1:
Master Equations and Fokker-Planck Equations}, Springer-Verlag, Berlin, 1999;
H. J. Carmichael, \textit{Statistical Methods in Quantum Optics 2:
Non-Classical Fields}, (Springer-Verlag, Berlin, 2008).

\bibitem {3}Wolfgang P. Schleich, \textit{Quantum Optics in Phase Space},
(Wiley-VCH, Birlin, 2001).

\bibitem {4}M. Hillery, R. F. O'Connell, M. O. Scully and E. P. Wigner, Phys.
Rep. \textbf{106,} (1984) 121.

\bibitem {5}Memorial Issue for H. Umezawa, Int. J. Mod. Phys. B \textbf{10},
(1996) 1695 memorial issue and references therein.

\bibitem {6}H. Umezawa, \textit{Advanced Field Theory -- Micro, Macro, and
Thermal Physics} (AIP 1993)

\bibitem {7}Y. Takahashi and H. Umezawa, Collecive Phenomena \textbf{2},
(1975) 55.

\bibitem {8}Hong-yi Fan and Yue Fan, Phys. Lett. A \textbf{246}, (1998) 242;
ibid, \textbf{282,} (2001) 269.

\bibitem {9}Hong-yi Fan and Yue Fan, J. Phys. A \textbf{35,} (2002) 6873;
Hong-yi Fan and Hai-liang Lu, Mod. Phys. Lett. B, \textbf{21,} (2007) 183.

\bibitem {10}Hong-yi Fan, Hai-liang Lu and Yue Fan, Ann. Phys\emph{.}
\textbf{321,} (2006) 480.

\bibitem {11}Hong-yi Fan, H. R. Zaidi and J. R. Klauder, Phys. Rev. D
\textbf{35}, (1987) 1831.

\bibitem {12}A. W\"{u}nsche, J. Opt. B: Quantum Semiclass. Opt. \textbf{1,}
(1999) R11.

\bibitem {13}E. P. Wigner, Phys. Rev. \textbf{40,} (1932) 749

\bibitem {14}G. S. Agarwal and E. Wolf, Phys. Rev. D \textbf{2,} (1970) 2161;
R. F. O'Connell and E. P. Wigner, Phys. Lett. A \textbf{83,} (1981) 145.

\bibitem {15}A. W\"{u}nsche, \textit{J. Computational and Appl. Math.}
\textbf{133} (2001) 665.

\bibitem {16}A. W\"{u}nsche, \textit{J . Phys. A: Math. and Gen. }\textbf{33}
(2000) 1603.

\bibitem {17}R. R. Puri, \textit{Mathematical Methods of Quantum Optics}
(Springer-Verlag, Berlin, 2001), Appendix A.

\bibitem {18}R. J. Glauber, Phys. Rev. \textbf{130,} (1963) 2529; Phys. Rev.
\textbf{131,} (1963) 2766.

\bibitem {19}J. R. Klauder and B. S. Skargerstam\textit{, Coherent States},
(World Scientific, Singapore, 1985).

\bibitem {20}C. Gardiner and P. Zoller, \textit{Quantum Noise} (Springer
Berlin, 2000).

\bibitem {21}Hong-yi Fan and Li-yun Hu, Opt. Commun. \textbf{282,} (2009) 932;
\textbf{281,} (2008) 5571.

\bibitem {22}G. S. Agarwal and K. Tara, Phys. Rev. A \textbf{43,} (1991) 492.

\bibitem {23}Li-yun Hu and Hong-yi Fan, Phys. Scr. \textbf{79,} (2009) 035004.

\bibitem {24}Hong-yi Fan and Li-yun Hu, Opt. Lett. \textbf{33,} (2008) 443.
\end{thebibliography}
\end{document}